\documentclass[prl,twocolumn,showpacs,floatfix,titlepage,superscriptaddress]{revtex4}

\usepackage{bm}
\usepackage{graphics} 
\usepackage{graphicx}  
\usepackage{latexsym}                
\usepackage{amsmath}     
\usepackage{amsfonts}  
\usepackage{amssymb}
\usepackage{amsthm}
\usepackage{dcolumn}
\usepackage{color}

\renewcommand{\phi}{\varphi}
\renewcommand{\>}{\right \rangle}
\newcommand{\<}{\left \langle}
\newcommand{\ket}[1]{\left |#1\>}
\newcommand{\bra}[1]{\<#1\right |}
\newcommand{\be}{\begin{equation}}
\newcommand{\ee}{\end{equation}}
\newcommand{\bea}{\begin{eqnarray}}
\newcommand{\eea}{\end{eqnarray}}

\begin{document}
\title{Route to Direct Multiphoton Multiple Ionization}

\author{P.~Lambropoulos}
\affiliation{Institute of Electronic Structure \& Laser, FORTH, P.O.Box 1527, GR-71110 Heraklion, Greece}
\affiliation{Department of Physics, University of Crete, P.O. Box 2208, GR-71003 Heraklion, Crete, Greece}
\affiliation{Kavli Institute for Theoretical Physics, Santa Barbara, California 93106, USA}

\author{G.~M.~Nikolopoulos}
\affiliation{Institute of Electronic Structure \& Laser, FORTH, P.O.Box 1527, GR-71110 Heraklion, Greece}
\affiliation{Kavli Institute for Theoretical Physics, Santa Barbara, California 93106, USA}

\author{K.~G.~Papamihail}
\affiliation{Department of Physics, University of Crete, P.O. Box 2208, GR-71003 Heraklion, Crete, Greece}

\date{\today}

\begin{abstract}
We address the concept of direct multiphoton multiple
ionization in atoms exposed to intense, short wavelength
radiation and explore the conditions under which such processes
dominate over the sequential. Their contribution is shown to be quite
robust, even under intensity fluctuations and interaction volume
integration, and reasonable agreement with experimental data is also
found.
\end{abstract}


\maketitle

The quest for direct multiple excitation/ionization of several electrons bound in atoms or molecules under the influence 
of intense laser radiation dates back to the early 80's \cite{Luk83,BoyRho85,PL85,PL87,Rob86}. 
Until the recent appearance of the FEL-based intense 
XUV and X-ray sources \cite{Ack07}, however, the hitherto available infrared and optical sources, although quite intense, proved 
inadequate for the task.  Multiple ionization has of course been observed in a number of experiments \cite{Luk83,BoyRho85,PL85,PL87,Rob86}, but the mechanism has been the sequential stripping of electrons, beginning with the valence shell and proceeding inwards. 

The only notable exception to the sequential stripping, under long wavelength radiation ($\sim 800$nm), came about with the advent of 
ultrashort  (subpicosecond) TiSa sources, which led to the observation of non-sequential double  ionization \cite{comment}.  
Still, the basic mechanism relies on a valence electron pulled out by the field, set into oscillatory motion thereby causing the ejection of a second, or perhaps third, electron by 
collision, as it returns towards the core; hence the term recollision \cite{comment}.  For this to be possible, the ponderomotive energy (cycle-averaged kinetic energy)
of a quasifree electron under the field must be larger than the binding energy of a second electron. 

The situation has now changed dramatically with the appearance of XUV to X-ray intense sources which has made feasible 
for the first time the observation of a number of non-linear processes in that wavelength range \cite{richter,mey10,glo10,Young10}.  
The decisive developments in that respect are the large peak intensity and sub-picosecond pulse duration. The latter is of 
central importance to our considerations in this work. An early, small scale so to speak, development in that direction has been 
the direct 2-photon double ionization of Helium which has grown into a subfield with tens of theoretical \cite{foumouo} and a few 
experimental studies \cite{mosch}, limited mainly by the present early stage of the sources.  The chief difference between this 
process and its counterpart under long wavelength is that no recollision is involved, as the ponderomotive energy is totally 
negligible. The two electrons are pulled out by the field while, electron-electron interaction although present, is not necessary; in contrast to single-photon double ionization. The information on this process, collected so far, can 
serve as a calibration for the larger scale generalization proposed in this paper.

Although the idea, under suitable conditions, is applicable to essentially any atom, in the interest of providing a 
quantitative assessment of the underlying physics, we focus on the specific context of Neon under radiation of 
photon energy 93 eV ($\approx$ 13.3 nm). In fact, some experimental data have already appeared in the literature \cite{richter09}.  
For this photon energy, even at a peak intensity  $10^{18}\rm{W}/\rm{cm}^2$, the ponderomotive energy 
$U_{\rm p}$ is about 10 eV, which is much smaller than both the photon energy and the binding energy of any electron in
Neon, thus guaranteeing the validity of LOPT (Lowest non-vanishing Order of Perturbation Theory) \cite{PL85,PL87,makris09}. 
In addition, for the notion of the cross section (generalized cross section of the appropriate order) to be valid, 
the pulse duration must be at least 10  cycles of the field. For photon frequencies of the order of 90-100 eV, even a pulse duration 
of 1fs amply satisfies this condition.  If the data are limited only to populations of the ionic species produced in the process, 
a set of kinetic (rate) equations are sufficient for the interpretation, as well as for our purpose in this paper.  
This is a set of differential equations governing the evolution of the 
populations of the various ionic species during the pulse \cite{PL87,makris09}. 
The complete set of such equations involving sequential, as well as direct processes from the neutral, 
leading to the ejection of up to 8 electrons are of the form \cite{suppl}
\begin{subequations}
\label{eqs}
\bea
\frac{dN_0}{dt} &=& 
-\sum_{j=1}^{8}\sigma_{0,j}^{(n_{0,j})} F^{n_{0,j}} N_0\\
\frac{dN_j}{dt} &=& 
\sigma_{0,j}^{(n_{0,j})} F^{n_{0,j}} N_0
+\sigma_{j-1,j}^{(n_{j-1,j})} F^{n_{j-1,j}} N_{j-1}\Theta[j-2]\nonumber\\
&-&\sigma_{j,j+1}^{(n_{j,j+1})} F^{n_{j,j+1}} N_j\Theta[7-j]
\eea
\end{subequations}
with $1\leq j\leq 8$ and $\Theta[j]$ the discrete Heaviside function.

The terms in the right hand side represent various processes contributing to the rate of change of the species whose time 
derivative appears on the left hand side. Thus $N_j$  indicates the $j$th 
ionic species of charge $(+j)$, while a term like $\sigma_{j,k}^{(n)} F^{n} N_{j}$ represents 
an $n$-photon process leading from species $j$ to species $k$, 
with the corresponding $n$-photon (generalized) cross section  $\sigma_{j,k}^{(n)}$, 
where $F(t)$ is the time-dependent photon flux in photons/cm$^2$sec. 
These equations are to be solved under a pulse, as realistic as possible, 
dictated by the conditions of operation of the source; in this case the FEL.

\begin{figure}[htp]
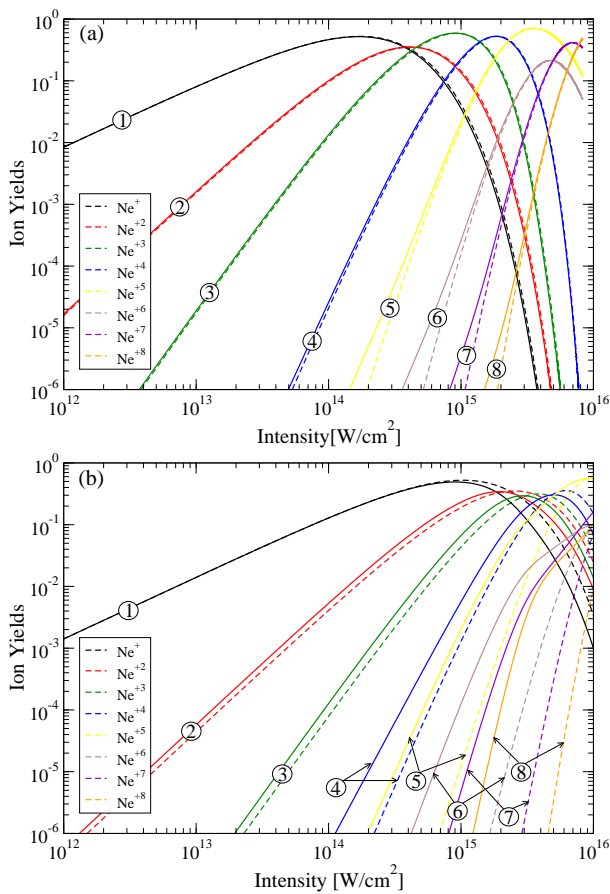

\centering
\includegraphics[width=8.cm,clip]{Deterministic30fs_NewCS-2_v2.eps}
\hfill
\includegraphics[width=8.cm,clip]{Deterministic5fs_NewCS-2_v2.eps}\\
\caption{(Color online) Ionization of Ne at 93 eV under a deterministic pulse 
with duration (a) 30 fs and (b) 5 fs. Solution of 
Eqs. (\ref{eqs}) in the presence of sequential channels alone (dashed lines)  
and with both direct and sequential processes included (solid lines).   
} \label{plot1.fig}
\end{figure}

The simplest scenario consists of retaining only sequential processes in 
the equations, which means that only the sequence of channels leading from ion $j$ to $j+1$, 
by successive single electron ejections, is retained in the equations. 
At relatively low peak intensity and long pulse duration, these will be the dominant channels. 
As a point of calibration for things to follow, we present in Fig. \ref{plot1.fig} 
the ion yields as a function of peak intensity, at the end of deterministic  
pulses of duration 30 fs [Fig. \ref{plot1.fig}(a)] and 5 fs [Fig. \ref{plot1.fig}(b)]. 
In both figures, the dashed lines represent the yields for the sequential channels alone. 
These figures depict the typical single atom behavior, illustrating the appearance 
and disappearance of ionic species as they give rise to higher ones with rising intensity.

For Ne under 93 eV photons, however, an entirely new class of channels
are energetically possible. These are direct, several electron multiphoton
(of the appropriate order) processes, leading from the neutral to the
corresponding ion. Specifically, 2 photons can eject 2 electrons leading to
Ne$^{+2}$, 3 photons can eject 3 electrons leading to Ne$^{+3}$, etc., up to 6
photons leading directly to Ne$^{+6}$. These are higher order generalizations
of 2-photon 2-electron ejection in He \cite{foumouo}. 
Moreover, we have included 
an 8-photon 7-electron transition Ne$\to$ Ne$^{+7}$, and an 11-photon 8-electron
transition Ne$\to$ Ne$^{+8}$. Direct $n$-photon $m$-electron
ejection can in principle always occur, for $n\geq m$, as long as it is
energetically allowed. It can in fact occur, from any ionic 
species, during
the interaction, but it is mostly from the neutral, where all of the
population resides at $t=0$, that such processes are expected to play 
an important role.
In our equations above, those processes are represented by the terms 
in the right hand side which contain the population $N_0$ of the neutral, 
involving also multiphoton cross sections of the appropriate order.

The values for direct $n$-photon $m$-electron ejection cross sections represent new territory. 
There is, however, a basis for conjecturing reasonable values.  
We argue that the cross section for $n$-photon, $m$-electron ejection (with $m<n$) 
in a given atom, is of the same order of magnitude as an $n$-photon one-electron 
ejection, for the same photon energy \cite{suppl}. The reasoning here rests on the following 
properties of a generalized cross section: (a) The $n$-photon transition  matrix 
element from the ground state connecting to $n$ electrons in the continuum requires 
no correlation. (b) As a consequence, the summation over intermediate states is dominated, 
in both cases, by single electron matrix elements. (c) All intermediate states are in the 
continuum, with no intermediate resonances of any significance. 
E.g., for 2-photon, 2-electron ejection from an initial state $1s^2$, one would have 
\bea
\sum_{\bf k} \frac{\bra{{\bf k}^\prime p,{\bf k}p}\hat{\bf r}_1+\hat{\bf r}_2\ket{{\bf k}p;1s} \bra{{\bf k}p;1s}\hat{\bf r}_1+\hat{\bf r}_2\ket{1s^2}}{E_{{\bf k}p}-E_{1s^2}-\hbar\omega}, 
\eea
with $\hat{\bf r}_i$ the electronic coordinates, and the summation extending over all allowed 1-electron excited states. This is non-zero even for non-interacting electrons, reducing to the product  $\bra{{\bf k}^\prime p}\hat{\bf r}_2\ket{1s}\bra{{\bf k} p}\hat{\bf r}_1\ket{1s}$ of 1-electron matrix elements.
One example of applicability of 
this conjecture is found in the cross section of two-photon double ionization of Helium which 
turns out to be roughly equal to the two-photon one electron ejection \cite{NakNikPRA}; 
while in contrast the cross section for one-photon two-electron ejection is smaller than the 
one-photon one-electron ejection by a factor of about 50.  The reason of course has to do 
with the fact that ejecting $m$ electrons by $n$-photon absorption, for $n< m$, is impossible without correlation. 
A second example is 4-photon double ionization of Carbon \cite{PL87}. 
Adopting the above conjecture, we obtain single-photon and two-photon
cross sections through a calculation and higher order cross sections
through a procedure of scaling \cite{PL87,makris09}.

The result of a calculation including these direct higher non-linearity channels 
is shown by solid lines in Fig. \ref{plot1.fig}. Their presence does not alter 
significantly the ion yields for the 30 fs pulse, but for the 5 fs pulse it does increase rather dramatically the yields of the higher charge species, 
beginning with Ne$^{+4}$. There is a clear physical interpretation of these results. 
The effect of the direct channels is more pronounced for the shorter pulse, because the sequential 
channels do not have as much a chance to drain the neutral --- from which the direct channels originate --- 
as they have in the rising wing of the longer pulse. And the direct channels, being of higher 
non-linearity, will dominate only if exposed to higher intensity, provided there is 
still enough population left in the neutral; which is the case for the shorter pulse. 
For the same reason, the yields of lower charge ions (up to Ne$^{+4}$) are not affected much because, 
in their case, even the direct channels are of low order. 

\begin{figure}[htp]
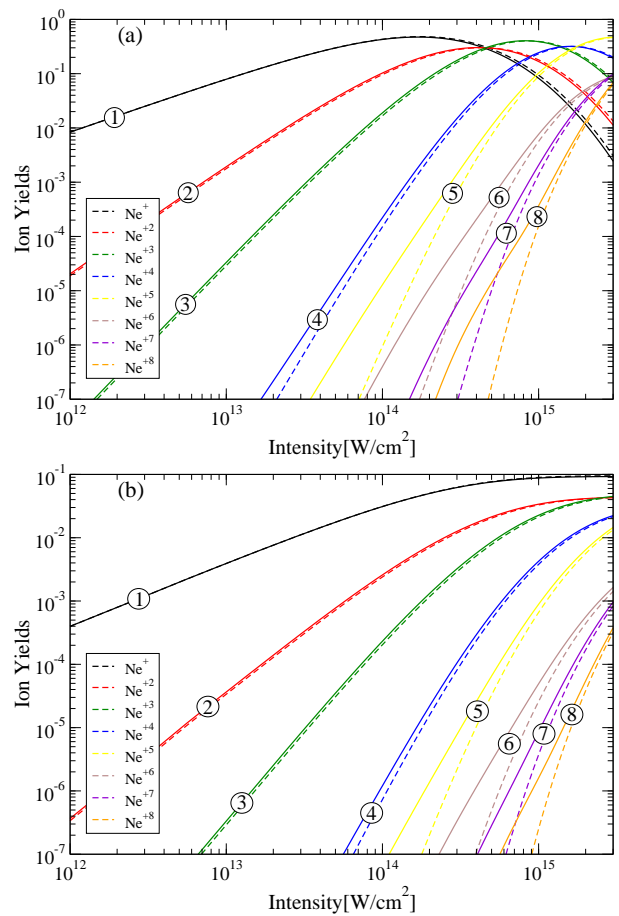

\centering
\includegraphics[width=8.cm,clip]{ChSD_vs_ChS30fs_Tc6fs_10000Trj_NewCS-2_v2.eps}
\hfill
\includegraphics[width=8.cm,clip]{VolumeIntChSD_vs_ChS30fs_Tc6fs_10000Trj_NewCS-2_v2.eps}\\
\caption{(Color online) Ionization of Ne at 93 eV under chaotic pulses of duration 30 fs and coherence time 6 fs. 
(a) Ion yields obtained by solving Eqs. (\ref{eqs}) for randomly chosen $F(t)$, 
in the presence of sequential (dashed) and sequential+direct (solid) channels. The presented 
yields are averaged over $10^4$ realizations.
(b) As in (a) with volume expansion effects included.   
} \label{plot2.fig}
\end{figure}

A first conclusion at this point is that, if the temporal shape of the pulse were deterministic, 
shortening its duration would enhance the contribution of the direct channels. 
It is, however, well known that, at least for the time being, 
the pulses of the FEL sources are not deterministic but exhibit strong intensity fluctuations. 
Qualitatively speaking, they can be considered chaotic \cite{Ack07}. 
Specifically, the overall pulse envelop is known to contain spikes 
of random height and duration, which can be as short as 5fs. 
This means that the atom is indeed exposed to spikes of duration sufficiently 
short to enhance the direct channels.
The implications for the theory is that Eqs. (\ref{eqs}) become stochastic, 
owing to the stochastic nature of the intensity.  
The simplest way to account for this, is to recall that $n$-photon ionization, 
within LOPT, 
is proportional to the $n$-th order intensity correlation function  $G^{(n)}$ \cite{LamAg}, 
which for a chaotic field is given by $n! \bar{F}^n$, where $\bar{F}(t)$ is the average 
intensity (flux). We could thus replace $F^n$ by $n! \bar{F}^n$ 
in Eqs.(1), which amounts to effectively increasing the 
$n$-photon ionization cross section by a factor of $n!$. 
Obviously, this favors processes of higher order.

This procedure would be rigorous if we had a single $n$-photon process and in addition 
the field were truly chaotic. What we have, however, is a set of differential equations 
coupling processes of various orders. Multiplying the cross sections by $n!$ and using a 
deterministic pulse amounts to a decorrelation approximation, valid if the 
ionic populations do not change appreciably on the scale of the intensity fluctuations. 
An alternative but rigorous approach is to introduce an appropriate stochastic model for the 
radiation, and solve the differential equations for a sufficiently large number of realizations of the radiation, 
taking in the end the average over such runs. This corresponds exactly to the manner experimental data are 
obtained.  By comparing the results with those obtained through the above mentioned decorrelation, 
we can also assess the limits of validity of the latter. Space does not allow a detailed 
comparison of the two models, nor a detailed description of our stochastic modeling.

In the following we focus on results obtained within the ab initio stochastic modeling 
of the field, relying on theoretical as well as experimental information about the FEL pulses 
(e.g., see \cite{Ack07,suppl,RohSan}).  
Comparing the averaged ion yields of Fig. \ref{plot2.fig}(a),  
to those of Fig. \ref{plot1.fig}(a), 
we note a dramatic increase of the higher charge species 
in the presence of chaotic light, when the direct channels are included.   
We have thus shown that intensity fluctuations will also enhance the direct channels over the sequential, 
even for the longer pulse of 30fs. 

\begin{figure}[htp]
\centering
\includegraphics[width=8.cm,clip]{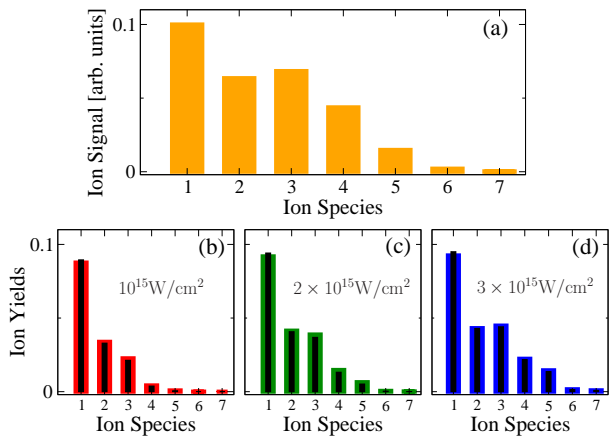}
\caption{(Color online)
Ionization of Ne under chaotic pulses of duration 30 fs and coherence time 6 fs. 
(a) Data obtained by \cite{richter09}.
(b-d) Ion yields at three different intensities, as obtained by Fig. \ref{plot2.fig}(b). 
The thin (black) bars correspond to sequential ionization whereas the broad (colored) bars 
refer to the case of open direct and sequential ionization channels.
} \label{plot3.fig}
\end{figure}

To compare with experimental data on Neon \cite{richter09}, we have performed an integration 
over the spatial distribution of the radiation in an interaction volume approximating the experimental arrangement. 
The result of a calculation for a 30 fs pulse, corresponding to that of the experiment, including 
fluctuations, is shown in Fig. \ref{plot2.fig}(b). As expected, the populations of the ionic species do not decrease beyond 
the saturation intensity, but exhibit a slow increase, due to the contribution of more atoms from the periphery 
of the interaction volume, where the intensity is lower than in the center. The crucial point, however, is that the 
contribution of the direct channels is found to dominate, for the higher species, even upon spatial integration. 
It is therefore evident that the effect of the multielectron direct channels is quite robust, as it survives practically 
intact, beyond the deterministic pulse.
Since only peaks of TOF (Time of Flight) results, at a single laser intensity, have been given in \cite{richter09}, we have estimated 
the relative magnitude of those peak heights and have plotted as a histogram the corresponding values 
in Fig. \ref{plot3.fig}(a). The histograms of Figs. \ref{plot3.fig}(b-d), correspond to the ion yields  
at three different intensities, at and around the nominal experimental one, as obtained from our Fig. \ref{plot2.fig}(b).  
The comparison with the experimental data demonstrates an overall reasonable agreement, 
especially for intensities $\sim 3\times 10^{15}\textrm{W/cm}^2$, 
but does also point to the 
sensitivity of the results to the value of the intensity. In the absence of experimental laser intensity dependences of 
the ionic species, we cannot offer at this time a more detailed evaluation of the agreement between theory and experiment.

We can nevertheless attempt to address the question of whether there are traces of the direct channels in 
the limited experimental data at our disposal. 
To this end, in Figs. \ref{plot3.fig}(b-d) we present the relative heights due to the sequential channels alone (narrow black bars) 
as well as the ones when both direct and sequential channels are present (broad colored bars). 
Depending on the exact intensity, the effect of the direct channels, although rather small, is noticeable particularly for 
higher ions in Figs. \ref{plot3.fig}(b,c). The reason for the small contribution of the direct channels is that the data have 
been obtained at or beyond the saturation intensity; not the optimal conditions for detecting the contribution of the direct, 
as clearly demonstrated in Figs. \ref{plot1.fig}(a)  and  \ref{plot2.fig}. Data over a range of intensities, below saturation, 
would provide a much clearer picture. Finally, somewhat surprising at first glance, for all three intensities the yield for the first ion (Ne$^+$) 
in the case of sequential ionization only, seems to be larger than the corresponding yield in the case of sequential+direct ionization. 
This, however, is consistent with the underlying physics, 
since the absence of direct channels leads to a much slower depletion of the neutral, from which direct channels originate, 
in favor of Ne$^+$. For all other ions, the situation is reversed for all three intensities.

In summary, we have introduced the idea of direct multiple ionization,
have explored its contribution in a specific realistic context, having
incorporated all relevant aspects, such as intensity fluctuations and spatial
radiation distribution. 
Reasonable agreement with
existing but limited experimental data has been found, highlighting at the
same time the need for more detailed data. The results in this work
provide a benchmark case and guide for the planning of future
experiments with the new FEL sources which are undergoing an
explosive expansion.

This research was supported by the EC RTN EMALI and in part by the National Science Foundation under Grant No. NSF PHY05-51164.
The authors also thank R. Santra for interesting discussions.

\end{document}